\documentstyle[prl,aps,epsf,multicol]{revtex}
\begin{document}
\draft

\def\i{\imath\,}
\def\ih{\frac{\imath}{2}\,}
\def\undertext#1{\vtop{\hbox{#1}\kern 1pt \hrule}}
\def\ra{\rightarrow}
\def\lfa{\leftarrow}
\def\ua{\uparrow}
\def\da{\downarrow}
\def\Ra{\Rightarrow}
\def\lra{\longrightarrow}
\def\ler{\leftrightarrow}
\def\lrb#1{\left(#1\right)}
\def\O#1{O\left(#1\right)}
\def\EV#1{\left\langle#1\right\rangle}
\def\tr{\hbox{tr}\,}
\def\trb#1{\tr\lrb{#1}}
\def\dd#1{\frac{d}{d#1}}
\def\dbyd#1#2{\frac{d#1}{d#2}}
\def\pp#1{\frac{\partial}{\partial#1}}
\def\pbyp#1#2{\frac{\partial#1}{\partial#2}} 
\def\pd#1{\partial_{#1}}
\def\br{\\ \nonumber & &}
\def\brr{\right. \\ \nonumber & &\left.}
\def\inv#1{\frac{1}{#1}}
\def\be{\begin{equation}}
\def\ee{\end{equation}}
\def\bea{\begin{eqnarray}}
\def\eea{\end{eqnarray}}
\def\ct#1{\cite{#1}}
\def\rf#1{(\ref{#1})}
\def\EXP#1{\exp\left(#1\right)} 
\def\INT#1#2{\int_{#1}^{#2}} 
\def\LHS{left-hand side }
\def\RHS{right-hand side }
\def\COM#1#2{\left\lbrack #1\,,\,#2\right\rbrack}
\def\AC#1#2{\left\lbrace #1\,,\,#2\right\rbrace}

\title{Fractionalization and confinement in the $U(1)$ and $Z_2$ 
gauge theories of strongly correlated systems}
\author{T. Senthil and Matthew P.A. Fisher}
\address{
Institute for Theoretical Physics, University of California,
Santa Barbara, CA 93106--4030
}

\date{\today}
\maketitle

\begin{abstract}
Recently,
we have elucidated the physics of electron fractionalization in 
strongly interacting electron systems using a $Z_2$ gauge theory formulation.  Here we discuss the connection  
with the earlier $U(1)$ gauge theory approaches based on the slave boson mean field theory. In particular,
we identify the relationship between the holons and Spinons of the slave-boson theory 
and the true physical excitations of the fractionalized phases that are readily described in the 
$Z_2$ approach.
 
\end{abstract}
\vspace{0.15cm}


\begin{multicols}{2}
\narrowtext

The tantalizing possibility that the electron is fractionalized ({\em i.e} broken apart)  
is a crucial ingredient of several qualitative or semi-quantitative pictures\cite{PWA,KRS,LN,z2short} of the 
physics of the 
cuprate high-$T_c$ materials. It is extremely important to have a firm theoretical
understanding of the physics of electron fractionalization before a complete theory based on 
such ideas can be developed for the cuprates. This paper addresses and resolves some 
theoretical issues that arise in this context.

A popular model Hamiltonian that is often used as the starting point of 
discussions of the physics of the cuprate materials is the $t-J$
model: 
\bea
\label{tJ}
H & = & H_0 + H_J  ,\\
H_0 & = & -t\sum_{<rr'>}\left(c^{\dagger}_{r \alpha} c_{r' \alpha} + h.c \right)
-\mu\sum_r N_r  ,\\
H_J & = & J\sum_{<rr'>} \left(\vec{S}_r \cdot \vec{S}_{r'} - \frac{1}{4}N_r
N_{r'} \right)  ,
\eea
with the constraint of no double occupancy of any site, {\em i.e}
$N_r = \sum_\alpha c^{\dagger}_{r \alpha} c_{r \alpha} \leq 1$. 
Here the $c_{r \alpha}$ are electron operators
at site $r$ with spin polarization
$\alpha$, and $\vec S_r = \frac{1}{2}c^{\dagger}_r \vec \sigma c_r$ is the 
usual spin
operator.

Equally popular, though not very succesful, are
attempts\cite{BZA,BA,oldgauge,LN}  to describe the possibility of electron
fractionalization in the phases  of
the $t-J$ model using the  ``slave'' boson representation
\be
c_{r\alpha} = h^{\dagger}_r s_{r\alpha} ,
\ee
where $h_r$ creates a hole at site $r$ and is bosonic (dubbed the ``holon''), and $s^\dagger_{r\alpha}$ creates a spinful 
fermion at 
site $r$ (dubbed the ``Spinon'' - note the uppercase `S'). It is assumed that the holon carries all the 
electrical charge while the Spinon carries all the spin of the electron. 
If the ``slave'' particles are liberated from
each other, fractionalization is achieved. 
There are several problems with this program, however, and despite
more than a decade of efforts it has not led to a clear description of 
fractionalized phases. 
In particular, as is well-known, 
the slave boson representation induces a 
local internal $U(1)$ gauge symmetry under which
\be
\label{int_u1}
h_r \ra h_r e^{i\theta_r};~~~s_{r \alpha} \ra s_{r \alpha} e^{i\theta_r} .
\ee
As a result there are strong interactions between the 
slave particles mediated by a compact $U(1)$ gauge field.
The properties of the resultant strongly coupled gauge theory
are extremely difficult to reliably analyse. It is generally believed\cite{RSSuN,CN}, however, that fluctuations of the
gauge field inevitably leads to a confinement of the 
slave particles  - effectively recombining them to form the electron.
This precludes any hopes of describing
fractionalized phases, within which the holons and Spinons can
propagate as independent excitations. 

In an alternate approach\cite{z2long}, we
have recently demonstrated that a general class of strongly 
interacting electron models can be 
recast in the form of a discrete $Z_2$ gauge theory.
This new formulation enables a 
definitive theoretical characterization of electron fractionalization.
Specifically, we demonstrated the possibility of obtaining fractionalized 
phases in two or higher spatial dimensions. In such a phase, the electron splits
into two independant excitations - the spin of the electron is carried by a
neutral fermionic excitation (the ``spinon'' - note the lowercase `s') and the
charge is carried by a bosonic excitation (the ``chargon''). There is a third
distinct excitation, namely the flux of the  $Z_2$ gauge field (dubbed the
``vison''). The vison is gapped in the fractionalized phase.  The $Z_2$ gauge
theory approach is closely related, and is indeed mathematically equivalent,
to the ideas on vortex pairing\cite{NLII} as a means to achieve
fractionalization in two spatial dimensions.

In view of the popularity of the slave boson
$U(1)$ gauge theory approach, it seems worthwhile to understand it's connections, if any, 
with the physics of the
fractionalized phases discussed in Ref. \cite{z2long}. In particular, a natural question 
is how the holon and Spinon operators introduced above are related to the 
physical excitations (chargons, spinons, and visons) of the fractionalized phases. 
In this paper, we will expose this connection. 

To this end, consider first the slave boson representation
of the $t-J$ model. The Hamiltonian Eqn.\ref{tJ} may readily be rewritten in terms of the 
operators $h_r$ and $s_r$, provided it is supplemented with the 
local constraints,
\be
\label{u1cstrt}
h^{\dagger}_rh_r + \sum_\alpha s^{\dagger}_{r \alpha} s_{r \alpha} = 1 ,
\ee
at each and every site of the lattice. 
The Hamiltonian is then invariant under (i) the global (electromagnetic)
$U(1)$ transformation $h_r \ra h_r e^{i\phi}$ (ii) global $SU(2)$ spin rotations $s_r \ra Us_r$
with $U \in SU(2)$ and (iii) the local internal gauge transformation Eqn. \ref{int_u1}.
Note that the constraint Eqn. \ref{u1cstrt} implies that the generator of the internal $U(1)$ 
gauge symmetry is fixed to be one at each lattice site. 

After expressing the partition function
as a functional integral, the theory proceeds by decoupling terms quartic in $h$ and $s$ with complex 
Hubbard-Stratonovich fields.
The resulting action takes the form
\be
S[h,s,a_0, \chi, \eta]  =  S_{\tau} + S_{\chi \eta} + S_{h} + S_{s} ,
\ee
\bea
& S_{\tau} &  =  \int_{\tau} \sum_r \bar{h}_r (\partial_{\tau} -ia_{0r})h_r +
\mu\bar{h}_rh_r + \bar{s}_{r\alpha} (\partial_{\tau} - ia_{0r})s_{r\alpha}
\nonumber \\ 
& S_{\chi \eta}  & =  2J\int_{\tau} \sum_{<rr'>}
\left(|\chi_{rr'}|^2 + |\eta_{rr'}|^2\right) , \nonumber \\  
& S_h  & =  -t\int_{\tau} \sum_{<rr'>}
\chi_{rr'}\bar{h}_r h_{r'} + c.c + O(h^4) , \nonumber \\ 
& S_s  & = - J \int_{\tau}
\sum_{<rr'>} \chi_{rr'}\bar{s}_rs_{r'} - \eta_{rr'}(s_{r\ua}s_{r'\da}
-s_{r\da}s_{r'\ua}) + c.c. .
  \nonumber
\eea
Here $\chi_{rr'}(\tau), \eta_{rr'}(\tau)$ are complex decoupling fields
living on the links of the spatial lattice, and the field
$a_{0r}$ is a Lagrange multiplier imposing the on-site constraint in Eqn. \ref{u1cstrt}. 

The properties of this action have been examined\cite{KL} within a mean-field approximation. Of
particular interest to us is the mean-field solution found when the deviation $x$ from half-filling is not too 
large: $<\chi_{rr'}> =\chi_0$, $<\eta_{rr'}> = \Delta_0 \alpha_{rr'}$ with $\chi_0$ and $\Delta_0$ being
real constants, and $\alpha_{rr'} = +1$ on horizontal bonds and $-1$ on vertical bonds. 
Within this mean field solution, the holons propagate freely with hopping 
$t_h = t \chi_0$. The Spinons also propagate
freely but are paired into a $d_{x^2 -y^2}$ condensate. It is hoped that this mean field solution
correctly describes the ``$d$-wave RVB'' state proposed pictorially by Anderson\cite{PWA} and 
Kivelson et.al.\cite{KRS}, provided the holons are not condensed.
This mean field state also has, at first sight, 
several properties in common with the nodal liquid 
state\cite{NLI,NLII,z2long}. The electron has apparently been broken apart into the holon and the Spinon.  

The crucial conceptual question is whether the fluctuations about the mean field solution
invalidate it's qualitative features. To discuss the fluctuations, note that the mean field 
solution breaks the internal $U(1)$ gauge symmetry. It is important therefore to keep two kinds
of fluctuations: (i) Fluctuations in the phase of $\chi$ - these become the spatial components
of a $U(1)$ gauge field\cite{RSSuN} and (ii) fluctuations in the phase of the 
Spinon pair condensate\cite{DHL}. To capture
the latter, we introduce a $d$-wave Spinon pair field $e^{i\varphi^{sp}_r}$ that couples to a 
$d_{x^2 - y^2}$ pair of Spinons centered at site $r$. 
Upon returning to a Hamiltonian description,
the fluctuations about the mean field state 
can then be described in terms of a simple effective Hamiltonian:
\bea
H_{eff} & = & H_{hol} + H_{s} + H_{pair} , \\
H_{hol} & = & -t_h \sum_{<rr'>} e^{ia_{rr'}} h^{\dagger}_rh_{r'}  + h.c.
+\mu\sum_r n_{hr} ,\\
H_s & = & -t_s \sum_{<rr'>} e^{ia_{rr'}} s_{r \alpha}^{\dagger} s_{r'\alpha} + h.c. , \\
H_{pair} & = & \Delta_0 \sum_r \left[ e^{i\varphi^{sp}_r}p_r + h.c\right] , \\ 
p_r & = & \sum^{\prime}_{r'}\alpha_{rr'}e^{ia_{rr'}}\left(s_{r\ua}s_{r'\da}
-s_{r\da}s_{r'\ua} \right) .
\eea
Here in the last equation $r'$ is nearest neighbour to $r$,
and $t_s = J \chi_0$.
For technical reasons we have used a number-phase representation of the holon operator:
$h_r = e^{i\phi_{hr}}$, with 
\begin{equation}
[\phi_{hr}, n_{hr'}] = i\delta_{rr'} ,
\end{equation}
where $n_{hr}$ is the 
holon number operator, corresponding physically to the hole density. 

As required, $H_{eff}$ is invariant 
under the 
internal $U(1)$ gauge transformation,
\bea
\label{int_u1'}
h_r \ra e^{i\theta_r}h_r,& ~~& s_{r\alpha} \ra e^{i\theta_r}s_{r\alpha} ,\\
a_{rr'} & \ra & a_{rr'} + (\theta_r - \theta_{r'}) ,\\
e^{i\varphi^{sp}_r} & \ra & e^{i\left(\varphi^{sp}_r + 2\theta_r\right)} .
\eea
As expected the Spinon pair field operator
$e^{i\varphi^{sp}_r}$ creates an excitation with two units of internal $U(1)$ gauge charge.
The operators $s_{r \alpha}$ create unpaired Spinons - the analog of BCS quasiparticles for this Spinon
pair condensate. Note that the number of unpaired Spinons $s^{\dagger}_{r \alpha} s_{r \alpha}$ is {\em not}
conserved.
  
The effective Hamiltonian above must be supplemented with the constraint in Eqn. \ref{u1cstrt}
that the generator of the internal $U(1)$ gauge transformation equal one at each lattice site. 
The total internal $U(1)$ gauge charge is given by 
\be
n_{int}(r) =n_{hr}+ 2n^{sp}_r + \sum_{\alpha}s^{\dagger}_{r\alpha}s_{r \alpha}
. \ee
Here we have defined a number operator for the Spinon pairs which satisfies $[\varphi^{sp}_r, n^{sp}_{r'}]
= i\delta_{rr'}$, and commutes with the $h,s$ operators. The constraint is therefore
\be
n_{int}(r) = 1 .
\ee

Our goal is to recast this Hamiltonian in terms of (physical) charge and spin operators 
which are invariant under the 
internal $U(1)$ gauge transformation - the physical motivation for doing so is that it is expected that 
only particles which do not carry this internal $U(1)$ charge are expected to survive the strong
confining effects of the interaction with the $U(1)$ gauge field. To that end, we follow closely the 
procedure introduced in Ref. \cite{z2long} to deal with {\em electron} Hamiltonians with structure similar to
that of $H_{eff}$ above. We first split
the Spinon pair creation operator into two equal pieces:
\bea
\left(b^{\dagger}_{sp,r}\right)^2 & \equiv & e^{i\varphi^{sp}_r} ,\\
b^{\dagger}_{sp,r} & = & e_r e^{i\varphi^{sp}_r \over 2} \equiv e^{i\phi^{sp}_r} .
\eea
Here $e_r = \pm 1$, and $b^{\dagger}_{sp,r}$ creates {\em half} a Spinon pair. 
It is readily seen that 
the phase $\phi^{sp}_r$ is conjugate to $n_{int}(r)$:
\be
[\phi^{sp}_r, n_{int}(r')] = i\delta_{rr'} .
\ee
Note that the constraint $n_{int}(r) = 1$ implies that the conjugate phase $\phi^{sp}_r$
fluctuates wildly. In particular, it precludes any breaking of the 
internal $U(1)$ gauge symmetry.
 
We next define new operators invariant under the internal $U(1)$ gauge transformation by 
binding half the Spinon pair to the holon and the Spinon:
\be
h_r = b^{sp}_r b^{\dagger}_r;~~s_r = b^{sp}_rf_r .
\ee
We will denote the operator $b_r$ the chargon\cite{Subir}, and the operator $f_r$ the spinon (note the lowercase `s').
The reason for this terminology is that, as we show below, these correspond precisely to the 
operators with the same names introduced in Ref. \cite{z2long}, and are indeed the physical excitations
in the fractionalized phase. 
Writing the chargon operator as $b_r = e^{i\phi_r}$, we see that the phase $\phi_r$ is conjugate to the 
hole density $n_{hr}$: $[\phi_r, n_{hr}] = -i$\cite{note}. 
Further the spinon number $f^{\dagger}_{r \alpha} f_{r \alpha}$ is equal to the number of 
unpaired Spinons $s^{\dagger}_{r \alpha} s_{r \alpha}$. 

Our plan is to now make a change of variables in $H_{eff}$,
trading in the holon ($h$), Spinon ($s_\alpha$) and Spinon pair operators
($e^{i\varphi^{sp}}$) in favor of the chargon ($b$), the spinon
($f_\alpha$) and the half-Spinon pair operators ($e^{i \phi^{sp}}$). 
It is important to emphasize that of these three new operators,
both the chargon and the spinon are invariant under the 
internal $U(1)$ gauge symmetry - and {\em all} the internal $U(1)$ charge is carried by the operator
$b^{sp}_r$. 
However, as elucidated in our earlier work\cite{z2long},
the very process of splitting the Spinon pair operator 
into two pieces introduces a {\it new} gauge symmetry - a $Z_2$ gauge symmetry.  Specifically,
the Hamiltonian $H_{eff}$, when re-expressed in terms of the three new operators, is invariant under the local transformation:
\be
b^{sp}_r \ra - b^{sp}_r;~~b_r \ra -b_r;~~f_r \ra -f_r  ,
\ee
at any given site $r$.  Moreover, it is necessary to impose
a (new) constraint\cite{z2long}
on the Hilbert space of these three new operators,
so that there is a one-to-one correspondence with the
Hilbert space of the original three operators.
The precise form of the new constraint in this case is,
\be
\label{z2cstrt}
n_{int}(r) -n_{hr} - f^{\dagger}_{r \alpha} f_{r \alpha} = even  .
\ee
Note also that
the hole density $n_{hr} = 1- N_r$ where $N_r$,  the total electrical charge at site $r$, is conjugate to the
chargon phase:  $[\phi_r, N_{r'}] = i \delta_{rr'}$. 
Thus, upon using the earlier constraint condition $n_{int}(r) = 1$, Eqn. \ref{z2cstrt} reduces to
\be
N_r - f^{\dagger}_{r \alpha} f_{r \alpha} = even .
\ee
This is exactly the same constraint on the chargon and spinon
numbers as in Ref. \cite{z2long}. 

Thus, we may now readily obtain a  
path-integral expression
for the partition function of $H_{eff}$ (in
terms of $b^{sp},b$ and $f_\alpha$) exactly as in 
Ref. \cite{z2long}. The $Z_2$ constraint above can be implemented by means of a projection
operator ${\cal P}_r = \frac{1}{2} \left(1 + (-1)^{N_r - f^{\dagger}_rf_r} \right)$ at each
lattice site. The $U(1)$ constraint $n_{int}(r) = 1$ is more conveniently implemented
by adding to $H_{eff}$ the term $U \sum_r (n_{int}(r) - 1)^2$ and 
letting $U$ go to infinity.  
As detailed in Ref. \cite{z2long}, the final result may essentially
be written down on symmetry grounds - which in the present context
are the 
$U_{int}(1) \times Z_2$ gauge symmetries, in addition to
the global charge $U(1)$ and spin $SU(2)$ symmetries.
The final action for the path-integral
(for large finite $U$) takes the form,
\bea
S & = & S_{z2g} + S_{sp} ,\\
S_{z2g} & = & S_c + S_s + S_B ,\\
S_{sp} & = & -\sum_{<ij>} t^{ij}_{sp} \sigma_{ij} cos(\phi^{sp}_i - \phi^{sp}_j + a_{ij}) .
\eea
Here $S_{z2g}$ is exactly the $Z_2$ gauge theory action in Ref. \cite{z2long}, and describes
chargons and spinons minimally coupled to a fluctuating gauge field $\sigma_{ij}$. 
(The indices $i,j$ label the sites of a space-time lattice). 
Note that all the coupling to the internal $U(1)$ gauge field is through the field
$\phi^{sp}_i$ as expected. The $Z_2$ gauge field $\sigma_{ij}$ couples together the $\phi^{sp}$
with the chargons and the spinons. But note that we may absorb the $\sigma_{ij}$ into the $a_{ij}$
by shifting,
\be
a_{ij} \ra a_{ij} + \frac{\pi}{2} (1 - \sigma_{ij}) .
\ee
Then the action for $\phi^{sp}$ simply becomes,
\be
S_{sp} = -\sum_{<ij>} t^{ij}_{sp} cos(\phi^{sp}_i - \phi^{sp}_j + a_{ij}) ,
\ee
and is then completely de-coupled from the $Z_2$ gauge theory.
We may then
integrate out the $\phi_{sp}$ and $a_{ij}$ fields without affecting the rest of the action, and this may be done for any $U$
including the limit $U \ra \infty$. 
We thereby obtain 
the $Z_2$ gauge theory of Ref. \cite{z2long},
consisting of chargon and spinons minimally coupled to the
$Z_2$ gauge field.

Having derived the $Z_2$ gauge action from the theory of fluctuations about 
the slave boson mean field theory,
we may now directly take over the discussion of fractionalization 
from Ref. \cite{z2long}. In particular, 
it is the chargon and spinon (lower case)
fields ($b$ and $f_\alpha$) which create the physical excitations 
in a fractionalized phase, and not the 
holons and Spinons of the slave boson theory. The latter carry 
an internal $U(1)$ gauge charge, and so 
are susceptible to the confining effects of the compact $U(1)$ gauge field. 
The chargons (spinons)
are obtained from these by binding half a Spinon pair to the 
holon (Spinon) thereby neutralizing their
internal $U(1)$ charge.  Instead the chargons and spinons are 
coupled to a $Z_2$ gauge field which allows them
to be deconfined in two or higher spatial dimensions\cite{z2long}.

The discussion in this paper can also be directly taken over to clarify some puzzling
cryptic remarks in the literature on the possibility of deconfined spin $1/2$ excitations in 
Heisenberg spin models\cite{RSSpN,Wen}. These works start with, for instance,  
the Schwinger boson representation
of the Heisenberg spins (which introduces a $U(1)$ gauge symmetry), and propose 
obtaining fractionalized phases by condensing pairs of Spinons, 
thereby reducing the gauge symmetry down to 
$Z_2$. However, with the constraint that the number of 
Schwinger bosons at each site is fixed, it would seem, at first sight, 
that breaking of the $U(1)$ gauge symmetry is prohibited. 
Nevertheless, the construction given in this paper
shows how one might get fractionalization without actually 
breaking the $U(1)$ gauge symmetry. The resulting fractionalized 
phases are then identical to those obtained by imagining that the Spinon 
pair field has condensed\cite{FrSh}.

Before concluding, we emphasize some important differences 
between the physical pictures 
for the under-doped cuprates suggested by the 
$U(1)$\cite{BZA,LN} and $Z_2$\cite{z2short} gauge theory 
approaches. 
In the $U(1)$ approach, the pseudogap line is associated with the pairing of Spinons. 
The $U(1)$ theory is, at present, not powerful enough to unambiguously identify
the true physical excitations of the  system below this temperature scale as a
result of the strong interactions with the $U(1)$ gauge field. 
By contrast, in the $Z_2$
approach the spinons (which {\em are} physical excitations in a
fractionalized phase) are {\em always} paired. The pseudogap line is
associated with the gapping out of the vison excitations\cite{z2short} - 
the vortex-like excitations in the $Z_2$ gauge field.  Once
the visons are gapped out, the spinons and chargons are liberated from each
other and are the legitimate excitations  of the system. 

We thank Chetan Nayak for provoking us to think through the contents of this paper.
We particularly thank Yong-Baek Kim for his insistence that 
we publish these results, and for several useful comments,
and S. Sachdev  
for his
very constructive criticism and clarifying discussions
on Ref. \cite{RSSpN}. This research was generously
supported by the NSF  under Grants DMR-97-04005,
DMR95-28578
and PHY94-07194.

\end{multicols}

\begin{references}
\bibitem{note} The minus sign in the commutator is a consequence of the particular definition of the chargon
operator. It actually ensures that the commutator of the chargon phase with the physical electrical charge
$N_r = 1-n_{hr}$ has the usual sign.
\end{references}
\end{document}